IAC-21-B2.2

# Development of a miniaturized laser-communication terminal for small satellites

Alberto Carrasco-Casado[a]*, Koichi Shiratama[a], Phuc V. Trinh[a], Dimitar Kolev[a],
Yasushi Munemasa[a], Yoshihiko Saito[a], Hiroyuki Tsuji[a], Morio Toyoshima[b]

[a] *Space Communication Systems Laboratory, National Institute of Information and Communications Technology,
4-2-1 Nukui-Kitamachi, Koganei, Tokyo, Japan 184-8795*, alberto@nict.go.jp
[b] *Wireless Networks Research Center, National Institute of Information and Communications Technology,
4-2-1 Nukui-Kitamachi, Koganei, Tokyo, Japan 184-8795*, morio@nict.go.jp

\* Corresponding Author

**Abstract**

Free-space optical communication is becoming a mature technology that has been demonstrated in space a number of times in the last few years. The Japanese National Institute of Information and Communications Technology (NICT) has carried out some of the most-significant in-orbit demonstrations over the last three decades. However, this technology has not reached a wide commercial adoption yet. For this reason, NICT is currently working towards the development of a miniaturized laser-communication terminal that can be installed in very-small satellites, while also compatible with a variety of other different platforms, meeting a wide span of bandwidth requirements. The strategy adopted in this design has been to create a versatile lasercom terminal that can operate in multiple scenarios and platforms without the need of extensive customization. This manuscript describes the current efforts in NICT towards the development of this terminal, and it shows the prototype that has been already developed for the preliminary tests, which are described as well. These tests will include the performance verification using drones first with the goal of installing the prototype on High-Altitude Platform Systems (HAPS) to carry out communication links between HAPS and ground, and later with the Geostationary (GEO) orbit, covering this way a wide range of operating conditions. For these tests, in the former case the counter terminal is a simple transmitter in the case of the drone, and a transportable ground station in the case of the HAPS; and in the latter case the counter terminal is the GEO satellite ETS-IX, foreseen to be launched by NICT in 2023.

## 1. Introduction

Free-space optical communications is a well-known solution with the potential to provide the massive bandwidth of optical-fiber technology, but without the requirements for physical cabling, allowing to greatly enhance the capabilities of wireless terminals. Specifically, space laser communications (lasercom) is becoming a mature technology and it has been demonstrated in orbit numerous times in the last few years [1]. The Japanese National Institute of Information and Communications Technology (NICT) is a pioneer in this research topic, and it has carried out some of the most-significant in-orbit demonstrations over the last three decades. Starting as early as 1994, NICT conducted the first demonstration of a space-to-ground communication downlink from the Japanese geostationary satellite ETS-VI [2]. After the world's first GEO-ground downlink with ETS-VI, NICT demonstrated the world's first LEO-ground downlink with OICETS together with JAXA in 2006 [3], and the first LEO-ground lasercom and quantum key distribution (QKD) experiments with a microsatellite (SOCRATES) in 2014 [4].

## 2. Strategy to adopt lasercom in real scenarios

Despite the many successful demonstrations that space lasercom has shown during these last decades in almost all imaginable scenarios and platforms, this technology has not reached a wide commercial adoption yet. At the same time, the communication requirements of wireless terminals have not stopped to increase, and mobile networks based on 5G (and beyond) will become one of the biggest business of the modern world as well as an enormous technical challenge to support billions of humans, things, vehicles, robots, etc. generating an unprecedented amount of data. In this context, lasercom is destined to play a key role. Therefore, closing the gap between research and application has become a must, and practical lasercom systems must be developed and deployed in real scenarios as soon as possible.

With this goal in mind, NICT has started developing a series of versatile lasercom terminals that can fit a variety of scenarios and platforms to cover the needs of different use cases, always related to high-speed communications where RF cannot be easily adopted. Fig. 1 shows several scenarios where NICT is currently planning to demonstrate lasercom links.

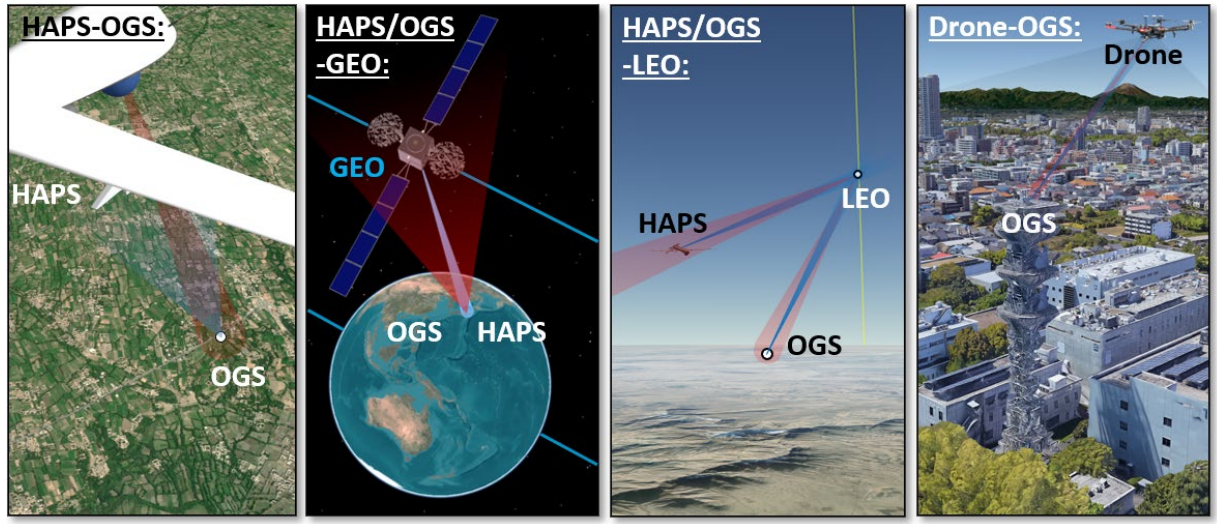

Fig. 1. Application examples of NICT's lasercom terminals in a variety of different scenarios and platforms

These scenarios are good examples of the requirements that lasercom terminals face if they must fit such different conditions without extensive customization, which should be avoided as much as possible to facilitate mass production and early adoption.

The first scenario shows an ultra-high speed bidirectional feeder link between a High-Altitude Platform System (HAPS) and a ground station to provide network access to a big number of users on the ground, generally based on RF connections (see Fig. 2). This could become a common use case to provide high-speed communications to a specific wide area because HAPS are like satellites from the point of view of the atmospheric propagation, but without the high cost of building and launching a satellite, and with a combination of the advantages of GEO satellites (being a stationary point in the sky) and LEO satellites (having a low latency), and additionally with a much more relaxed link budget due to the shorter distance.

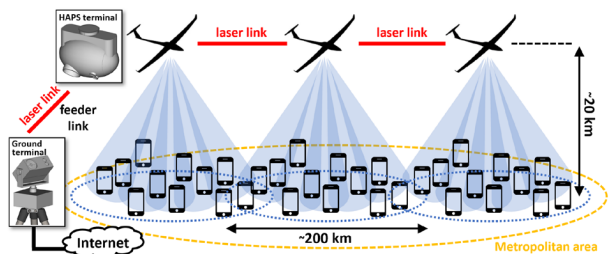

Fig. 2. Lasercom providing high-speed feeder link to support RF user connections on the ground

The second scenario is also a feeder link between the GEO orbit and ground or a terminal in the middle, for example onboard a HAPS or an airplane. The latter will be another application on the rise when high-speed Internet connections popularize on commercial flights. This is a worst-case scenario because something similar but easier in terms of link budget could be achieved by using a constellation of LEO or MEO satellites, which is a network configuration that is getting more and more popular. It is also the most technically challenging of these scenarios due to the long distance from the GEO orbit, and for this reason, the lasercom terminal described in this paper was designed to be able to support even this scenario as a worst case.

The third scenario includes LEO satellites, which is another platform that has been experiencing and will continue to experience an increasing demand, especially due to the lower cost of developing small satellites and CubeSats, as well as the launch cost. In particular, LEO constellation is a scheme that has re-emerged as a key part of the future communication networks as well as Earth-observation space infrastructure, with hundreds of satellites already in orbit as part of some existing constellation, and thousands of satellites planned to be launched in the coming years. Finally, the fourth scenario includes drones, which form another platform gaining popularity with big potential in a number of applications. For the two latter scenarios, a simplified version of the lasercom terminal is currently being designed and developed.

**3. Miniaturized lasercom terminal for small satellites**

In order to support the scenarios shown in the previous section, which covers a wide variety of concepts of operations as well as platforms with different requirements, a miniaturized lasercom terminal was conceived. The original specifications were chosen to meet the requirements of the most-challenging scenario and platform, namely, a LEO-to-GEO intersatellite link when the terminal is embarked in a CubeSat. The first conceptual design of this terminal, called CubeSOTA (CubeSat version of the NICT's SOTA, Small Optical TrAnsponder), was originally described in [5]. When considering the lasercom terminal together with its host satellite, the potential improvement compared to the previous NICT's LEO missions was estimated in a mass reduction of more than one order of magnitude and a datarate increase of three orders of magnitude in a time span of less than 20 years (Fig. 3).

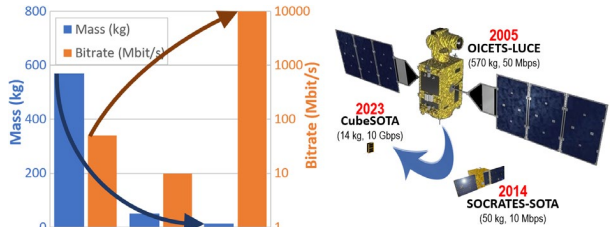

Fig. 3. Evolution of NICT's lasercom LEO satellites

Since the CubeSat form factor imposes the limitation of 10-cm cubes when designing the payload structure, the biggest practical telescope´s aperture that can be integrated in a CubeSat is a 10-cm diameter, but considering the structural support and physical implementation, a miniaturized telescope with an effective aperture of 9-cm was designed and developed with the minimum possible distance between primary and secondary mirror (Fig. 4). This miniaturized Cassegrain telescope has a 40× magnification to satisfy the size constraints of the internal beam, and the exit port is placed in one lateral side, where the rest of the optics are located.

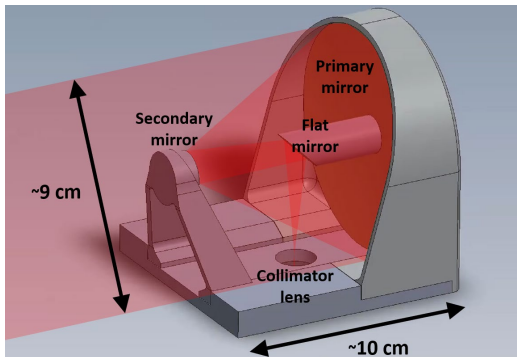

Fig. 4. 3D model of the 9-cm telescope

The 9-cm telescope is one of the most-advanced parts of the lasercom terminal at the time of writing this manuscript, and it is currently being tested for space qualification. Specifically, the wavefront error is being measured with a Fizeau interferometer during different stages of the thermal cycling to guarantee the required optical quality. The measured data shows a wavefront error of λ/19 RMS at 1550 nm and a total transmission of 93%, including all the elements of the telescope, i.e. primary mirror, secondary mirror, folded mirror, and collimator lens (see Fig. 4).

Another key component of a space lasercom terminal is the optical amplifier, which is necessary to achieve the required high transmitted power to close the link. The highest optical power that was determined to be feasible to achieve within the constrained volume of a CubeSat was 2 W. Based on this requirement, a miniaturized (95×95×25 mm) space-grade 2-stages Erbium-Doped Fiber Amplifier (EDFA) was designed and developed to be compatible with the CubeSat format factor (Fig. 5). This amplifier has been qualified for space environment after successfully gone through sinusoidal and random vibration tests, shock tests, thermal vacuum tests, and total ionization dose (TID) tests following the ISO 19683:2017 and ISO 15864:2004 standards.

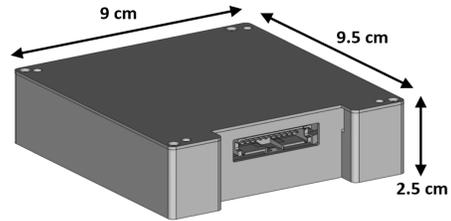

Fig. 5. 3D model and image of the CubeSat's EDFA

Fig. 6 shows an example of the dependence of the EDFA's output power on the case temperature during one hour of a thermal vacuum chamber (TVAC) test. As it can be seen, when the temperature rises, the output power drops. Identifying this trend, it is possible to guarantee the required output power depending on the initial temperature, the initial power, and the operation time. LEO-ground downlinks occur during passes of less than 10 minutes, and typically last less than 6 minutes [6]. Since this EDFA can start transmitting up to 2.5 W, the effective output power can be guaranteed to be over 2 W during the lasercom link. Therefore, this EDFA is suitable for LEO-ground communications. When considering links longer than one hour, additional heat transfer materials or method should be applied.

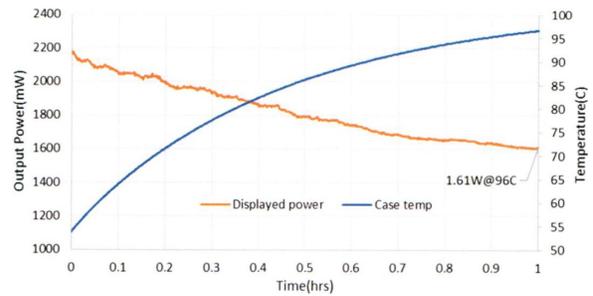

Fig. 6. EDFA's temperature vs power under TVAC test

Fig. 7 shows the 3D model of the first lasercom-terminal prototype. Although the terminal itself was initially designed to fit within the CubeSat form factor, a 2-axis gimbal was also developed to be able to operate it with other types of platforms. For example, when considering the demonstration with GEO, the EDFA requirements are too demanding for a CubeSat platform, and it was decided to carry out these first experiments using HAPS, for which a gimbal is necessary.

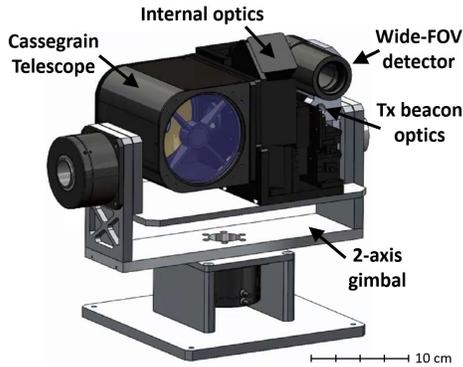

Fig. 7. 3D model of the first terminal prototype

Fig. 8 shows the first terminal prototype with all the basic functions that can be seen in the block diagram of Fig. 9, including closed loop between wide-FOV detector and 2-axis gimbal, closed loop between fine-tracking detector and fine-pointing mirror, point-ahead mechanism, 10-Gbps test modem, beacon transmitting system, and control unit to govern the terminal. Although this first prototype did not include dedicated optics for the receiver, this new function is currently being added to the terminal to support a receiver with single-mode fiber coupling.

This first prototype was designed and developed to be used in the experiments with the NICT's HICALI lasercom terminal onboard the ETS9 satellite, which will be launched in 2023 to the GEO orbit [7], as shown in the second illustration of Fig. 1. As explained before, the experiment with this terminal will be initially carried out using HAPS, and the first step will be to test HAPS-ground communications as shown in the first illustration of Fig. 1. Currently, this prototype is being adapted for HAPS, and the plan is to use NICT's transportable optical ground station to support the terminal on the ground [8]. The first verification phase will be done during 2022 with the terminal in the ground, and a simplified counterpart onboard a drone, and in a second phase mounting the terminal on the drone.

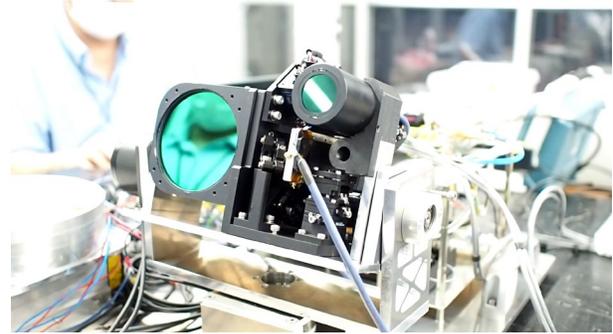

Fig. 8. Picture of the first terminal prototype

## 4. Conclusion

This paper presents the first prototype of NICT's miniaturized lasercom terminal for space applications, which design was done with the goal of having the potential to fit a variety of different scenarios and platforms while offering full high-speed bidirectional-communication capabilities up to long distances. The first prototype of the terminal was introduced and the development and test plan was described.

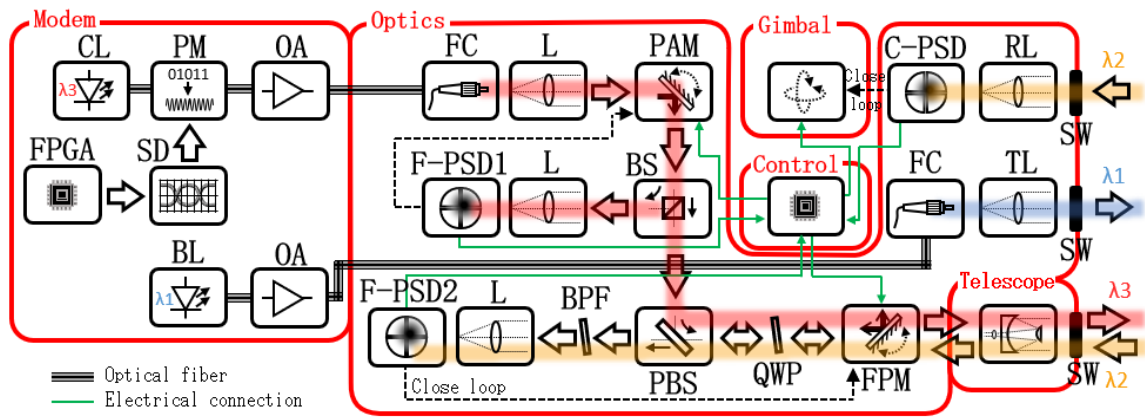

Fig. 9. Block diagram of the first prototype of the NICT's miniaturized lasercom-terminal